\documentclass[sigconf]{acmart}
\AtBeginDocument{%
  }

\setcopyright{acmlicensed}
\copyrightyear{2025}
\acmYear{2025}
\acmDOI{XXXXXXX.XXXXXXX}
\acmConference[WebSci '25]{ACM Web Science Conference 2025}{May 2025}{New Brunswick, New Jersey, USA}
\acmISBN{978-1-4503-XXXX-X/2025/05}

\begin{document}

\title{Evaluating Machine Expertise: How Graduate Students Develop Frameworks for Assessing GenAI Content}

\author{Celia Chen}
\email{clichen@umd.edu}
\orcid{0009-0008-9967-8968}
\affiliation{%
  \institution{University of Maryland}
  \city{College Park}
  \state{Maryland}
  \country{USA}
}

\author{Alex Leitch}
\email{aleitch1@umd.edu}
\orcid{INSERT HERE}
\affiliation{%
  \institution{University of Maryland}
  \city{College Park}
  \state{Maryland}
  \country{USA}
}

\begin{abstract}
This paper examines how graduate students develop frameworks for evaluating machine-generated expertise in web-based interactions with large language models (LLMs). Through a qualitative study combining surveys, LLM interaction transcripts, and in-depth interviews with 14 graduate students, we identify patterns in how these emerging professionals assess and engage with AI-generated content. Our findings reveal that students construct evaluation frameworks shaped by three main factors: professional identity, verification capabilities, and system navigation experience. Rather than uniformly accepting or rejecting LLM outputs, students protect domains central to their professional identities while delegating others—with managers preserving conceptual work, designers safeguarding creative processes, and programmers maintaining control over core technical expertise. These evaluation frameworks are further influenced by students' ability to verify different types of content and their experience navigating complex systems. This research contributes to web science by highlighting emerging human-genAI interaction patterns and suggesting how platforms might better support users in developing effective frameworks for evaluating machine-generated expertise signals in AI-mediated web environments.
\end{abstract}




\received{10 April 2025}
%%\received[accepted]{5 June 2009}

\maketitle

\section{Introduction}
The emergence of large language models (LLMs) presents an opportunity to examine how individuals evaluate and respond to machine-generated content that mimics human expertise across multiple domains \cite{bender2021}. As these systems become increasingly embedded in web-based platforms and educational contexts, understanding how users develop frameworks to assess their output is relevant for web science \cite{gillespie2020}. This paper focuses on graduate students, individuals who occupy a middle ground between novice and expert, and how they negotiate machine-generated expertise not only in academic work but also in personal, domestic, and emotional domains.

The recent proliferation of generative AI tools, such as ChatGPT, has altered the way information is accessed and processed on the Web \cite{manyika2019}. Unlike previous waves of technological innovation that primarily modified the material conditions of learning (such as smartboards and personal computing devices) \cite{ames2019, wood2008}, LLMs engage directly in domain-specific dialogue while lacking true understanding. As Chen and Leitch (2024) argue \cite{chen2024}, these tools appear to realize "a long-held dream of simulation: someone to talk to, who can talk back," creating systems that can provide patient, repetitive explanations during learning processes—work traditionally performed by teaching assistants, tutors, and peer mentors \cite{england2005}.

Graduate students represent a population through which to examine these interactions. As emerging professionals developing from novice to expert within their fields \cite{crider2024}, they must simultaneously develop their own domain expertise while learning to effectively evaluate and utilize tools that can generate seemingly authoritative responses \cite{collins2007}. Their position at this threshold makes their evaluation strategies relevant to understanding how expertise signals are interpreted in AI-mediated web environments \cite{eyal2013}.

This study addresses a central question with implications for web science: How do graduate students develop frameworks for evaluating the content of machine-generated responses? By examining strategies students employ to assess and engage with LLM-generated content, we gain insight into emerging behaviors that evolve through reliance on generative AI, behaviors that complement and contradict traditional expertise evaluation \cite{donath2007}. Our findings reveal patterns in how students protect domain-specific expertise while strategically delegating others \cite{abbott1988}, develop trust through verification strategies \cite{rader2018}, and transfer existing social signal interpretation abilities to novel technological contexts.

Understanding these frameworks is vital, as generative AI is becoming more difficult to avoid in educational and information-access contexts, particularly online. The findings of this study contribute to our understanding of human-genAI interactions on the web by revealing how users develop evaluation and engagement strategies for auto-generated content that mimics a tone of confident expertise.

\section{Methods}
To examine how graduate students develop frameworks for evaluating machine-generated content, we conducted a qualitative study combining multiple data sources across two semesters (Spring and Fall 2024).

\subsection{Participants}
Our study included 14 graduate students enrolled in required Human-Computer Interaction Master's (HCIM) courses at a large research university in the United States. Approximately two-thirds of participants were international students, primarily from India and China, with varying levels of English proficiency. The majority of participants were female, and they represented diverse professional backgrounds, from recent graduates to mid-career professionals returning to academia.

\subsection{Data Collection}
We collected data through three sources across the two semesters. The longitudinal survey component consisted of three instruments administered to each participant: a pre-study baseline assessment, a mid-semester evaluation, and an end-of-semester reflection. These surveys tracked changes in participants' attitudes toward and engagement with LLM systems, combining quantitative measures with open-response fields. All fourteen participants completed the full survey sequence.

For LLM interaction data, participants captured their course-related interactions through full-page screenshots, as there is no official export functionality in the LLM systems. Ten participants provided these interaction records, while four were unable to do so due to not having accounts with any of the LLMs they used.

Hour-long semi-structured interviews were conducted with each participant near the semester's end. Interviews explored how students developed frameworks for evaluating machine-generated content, focusing on their decision-making processes about when and how to engage with LLM systems.

\subsection{Data Analysis}
Interviews were recorded through Zoom and automatically transcribed using ATLAS.ti. Our analysis followed principles of thematic analysis \cite{braun2006} while drawing on established qualitative coding techniques \cite{charmaz2014, glaser1967}.

The analytical process began with initial open coding to identify patterns in how students evaluated and engaged with machine-generated content. First cycle coding remained close to participants' language and descriptions, generating codes like 'protecting core expertise' and 'verifying technical outputs.' Second cycle coding then grouped these initial codes into broader analytical categories.

Using an iterative process that included examining interview transcripts, survey responses, and interaction logs, we identified recurring patterns in self-described interactions and points of convergence and divergence between reported and observed behaviors. Theoretical saturation was assessed through the stability of core themes across later interviews and the diminishing returns in novel patterns from additional data collection.

\section{Findings}
Our analysis revealed patterns in how graduate students develop frameworks for evaluating and engaging with machine-generated content. We focus on three key frameworks that emerged from our data: (1) professional identity and expertise domain protection, (2) system trust development through verification strategies, and (3) cross-cultural navigation and adaptation.

\subsection{Professional Identity and Expertise Domain Protection}
Participants' decisions about which tasks to delegate to LLMs were strongly tied to their professional identities and how they understood signals of expertise in their chosen domains. Rather than making simple efficiency calculations, students carefully protected areas they considered central to their professional identity while strategically delegating others.

Students who identified primarily as managers readily delegated technical implementation while protecting work that signaled managerial expertise. One participant explained: "In terms of actually building that webpage and very technical details I could not claim credit for it but because I was taking the class and understanding what was being done and the concepts of it, I felt like I had my own personal coder to do the work for me sort of like I had a development team of one to execute the ideas I had."

In contrast, students with artistic backgrounds demonstrated strong protection of design-related expertise signals while delegating technical tasks: "I don't think I would ever use it for the creative part... With code I would go 'oh why didn't I know this, why didn't I study better' for me to write the code on my own... I don't even think I want to get super good at [coding], so it was fine when the LLM was really good at it. But creative work where I want to be able to say this work is my own at the end of the day I wouldn't do it because I really like that work."

Students who identified as programmers displayed a third pattern, protecting opportunities to demonstrate coding expertise while more readily delegating other tasks: "I make sure I have the time to approach something with an LLM, and I don't use it to push out a rush job... it's kind of an ego thing, I need to know I can approach the problem solo before producing LLM content."

These patterns suggest participants' understanding of which signals matter most in different professional contexts, and reflect what Abbott \cite{abbott1988} terms "jurisdictional claims" over knowledge domains - careful choices about which expertise signals to develop independently versus which can be safely delegated to AI systems.

\subsection{System Trust Development and Verification Strategies}
Participants demonstrated varying approaches to evaluating LLM expertise signals, with verification patterns emerging for different types of claimed expertise. For technical tasks, particularly coding, students expressed confidence in their ability to validate the system's expertise claims through direct assessment. However, they showed more hesitation with academic content where expertise signals were harder to verify.

One participant spoke explicitly on this distinction: "For Claude, I feel like with coding you can see that it works, like if you just enter it iteratively it starts to work now but if the academic text is so difficult I don't think Claude can do it... When I was reading more hardcore academic texts, I put some terms in Claude and I just couldn't trust it. For me, I just don't think of LLMs as smart enough; I mean I still asked it, but I felt like it just couldn't do it, explaining the hard academic texts to me."

This distinction in trust levels reflects an implicit understanding of different types of expertise signals. When students could directly verify the system's claimed expertise through what Donath \cite{donath2007} would term "assessment signals" - like code that either works or doesn't - they expressed more confidence in the system's capabilities. However, for tasks requiring deep comprehension or analysis, where expertise signals were harder to verify, students often expressed deeper skepticism about the system's claimed expertise.

These patterns suggest that students developed internal frameworks for evaluating machine-generated expertise signals, with their trust levels closely tied to their ability to independently verify the system's claimed competence in different domains.

\subsection{Cross-Cultural Navigation and Adaptive Strategies}
International students often developed strategies for using LLMs that built on their existing experience navigating unfamiliar institutional systems. Many international students developed structured verification strategies that leveraged LLMs across languages. As one international student described: "The Chinese version does really good now...I prefer the English first so I know the original terminology, then I use the Chinese to make sure I 100\% understand the words...Chinese is the double-check."

Students also demonstrated strategic approaches to managing academic demands, often drawing on their experience navigating institutional support systems: "I usually use these features in [one of] the reading response[s]...I was so tired, and I have something in my mind, but I don't know how to structure everything. So I just using the speech function and say, okay, I will tell you everything in my mind. Please structure that for me...in 300 words."

These patterns suggest that students who have developed strategies for navigating unfamiliar institutional systems often transfer these skills to LLM interaction. Their approaches reflect what Eyal \cite{eyal2013} describes as a deployment of various forms of capital, with students leveraging their existing system navigation expertise to evaluate and use machine-generated content for various academic tasks.

\section{Discussion and Implications}
This study reveals how graduate students develop frameworks for evaluating machine-generated content, offering several implications for understanding human-genAI interactions on the web.

\subsection{Value-Based Expertise Evaluation}
The patterns of value-based task delegation observed in this study reveal how users develop frameworks for evaluating different types of machine-generated expertise. Drawing on Collins and Evans' \cite{collins2007} distinction between contributory and interactional expertise, students' selective engagement with LLMs reflects mindful calculations about which domains of expertise they must develop through direct practice versus which can be supported through machine-generated content.

These patterns are relevant for web science as they highlight how users navigate what Searle's Chinese Room metaphor presents as a tension: systems that can produce apparently expert outputs without genuine understanding. Students' evaluation strategies suggest they implicitly recognize this distinction, developing frameworks for when pattern-matched outputs are sufficient versus when genuine understanding is necessary.

The strong reactions displayed toward LLM use in core professional domains - from artists protecting creative work to programmers insisting on independent problem-solving - reflect how expertise signals are transformed when moved into AI-mediated spaces. This finding has implications for how web platforms might design interfaces that acknowledge and support these domain-specific evaluation frameworks.

\subsection{Context-Dependent Verification Strategies}
Students developed verification strategies for different types of machine-generated content, suggesting that signal evaluation in digital spaces is not uniform but highly contextual. When students could directly verify the system's outputs through demonstrable success (e.g., code that runs correctly), they expressed more confidence in delegating those tasks. However, for domains requiring deeper comprehension where verification was harder to ensure, students often developed more elaborate frameworks for evaluating machine-generated content.

This finding extends our understanding of how expertise signals are assessed in AI-mediated web environments. Rather than treating all system outputs uniformly, users appear to develop layered approaches to verification, creating different standards to evaluate different types of machine-generated content. This suggests that web platforms incorporating generative AI should consider providing different types of supporting evidence or verification tools depending on the domain of expertise being conveyed.

\subsection{Accessibility and System Navigation}
The findings suggest that students' ability to effectively evaluate and leverage machine-generated content are connected to their existing frameworks for navigating complex systems. While LLMs may be technically available to all students, meaningful use appears to involve more than just access. Students' preexisting system navigation skills seem to influence how they approach evaluating machine-generated content, suggesting that tool availability alone may not address educational resource disparities.

This pattern highlights useful considerations for web science: as generative AI becomes more embedded in web-based information access, attention must be paid not just to technical availability but to how different user groups develop frameworks for evaluating and leveraging machine-generated content. Web platforms might consider how to scaffold evaluation skills for users with varying levels of system navigation expertise.

\section{Conclusion}
This study examined how graduate students develop frameworks for evaluating machine-generated content in academic and personal contexts, revealing distinct patterns in how expertise signals are assessed when traditional markers of authority are neither clearly present nor absent.

Students developed nuanced frameworks for evaluating different types of machine-generated content, from direct assessment of technical outputs to careful verification of academic content. These evaluation frameworks often drew on their existing abilities to interpret both professional and social signals, suggesting that expertise evaluation in digital spaces builds on rather than replaces traditional forms of signal interpretation.

These findings extend our understanding of human-genAI interactions on the web by revealing how users develop strategies for engaging with machine-generated content that mimics expertise signals without possessing genuine understanding. As generative AI becomes increasingly embedded in web-based information access and educational contexts, understanding these evaluation frameworks becomes crucial for developing systems that effectively support rather than undermine user agency.

Future research should examine how these frameworks evolve over time as users gain more experience with generative AI systems, and how they might vary across different disciplinary contexts and professional domains. Additionally, exploring how platforms might better support users in developing effective evaluation frameworks represents an important direction for web science research.

\bibliographystyle{ACM-Reference-Format}
\bibliography{workshop_ref}

\end{document}